1    **Title:** Starch Metabolism Has a Role in Juvenile-to-Adult Phase Transition in *Arabidopsis*

3    **Running Title:** Regulation of juvenility in *Arabidopsis*

5    **Authors:** Ianis G. Matsoukas, Andrea J. Massiah, Brian Thomas

7    **Authors' Addresses:** School of Life Sciences, Gibbet Hill Campus, The University of

8    Warwick, Coventry CV4 7AL. United Kingdom.

10   **Corresponding Author**: Ianis G. Matsoukas

11   Faculty of Advanced Engineering & Sciences, The University of Bolton, Deane Road, Bolton

12   BL3 5AB, United Kingdom.

13   Tel: +44 (0) 120 490 3409, Fax: +44 (0) 120 490 3088

14   Email: I.Matsoukas@bolton.ac.uk

1 **ABSTRACT**

3 The physiology and genetics underlying juvenility is poorly understood. Here we exploit

4 *Arabidopsis* as a system to understand the mechanisms that regulate floral incompetence during

5 juvenility. Using an experimental assay that allows the length of juvenility to be estimated, and

6 mutants impaired in different pathways, we show that multiple inputs influence juvenility.

7 Juvenile phase lengths of wild type (WT) accessions Col-0, L*er*-0 and Ws-4 are shown to differ,

8 with Col-0 having the shortest and Ws-4 the longest length. Plants defective in sugar signalling

9 [*gin1-1, gin2-1, gin6* (*abi4*)] and floral repressor mutants [*hst1, tfl1, tfl2* (*lhp1*)] showed

10 shortened juvenile phase lengths, compared to their respective WTs. Mutants defective in starch

11 anabolism (*adg1-1, pgm1*) and catabolism (*sex1, sex4, bam3*) showed prolonged juvenile phase

12 lengths compared to Col-0. Examination of diurnal metabolite changes in *adg1-1* and *sex1*

13 mutants indicates that their altered juvenile phase length may be due to lack of starch turnover,

14 which influences carbohydrate availability. In this article we propose a model in which a variety

15 of signals including floral activators and repressors modulate the juvenile-to-adult phase

16 transition. The role of carbohydrates may be in their capacity as nutrients, osmotic regulators,

17 signalling molecules and/ or through their interaction with phytohormonal networks.

1    **INTRODUCTION**

3    Plants undergo a series of qualitative transitions during their life cycle in response to both

4    environmental and endogenous cues. One of the most distinguishable is the transition from a

5    vegetative-to-reproductive phase of development. This stage is preceded by the juvenile-to-adult

6    phase transition within the vegetative phase. During the juvenile phase plants are incapable of

7    initiating reproductive development and are insensitive to environmental stimuli such as

8    photoperiod and vernalization, which induce flowering in adult plants. The juvenile-to-adult

9    phase transition has long attracted interest as an important developmental trait, especially in

10   those species where juvenility is prolonged. Knowledge gained about regulation of the juvenile-

11   to-adult phase transition could help with crop scheduling, decrease time to flowering, and reduce

12   waste with resulting benefits for the environment through lower inputs and energy required per

13   unit of marketable product.

14       The genetics and physiology underlying the juvenile-to-adult phase transition is poorly

15   understood. This transition may be associated with physiological, morphological and

16   biochemical markers (Thomas & Vince-Prue 1984; Poethig 1990). However, these changes are

17   often less distinct in herbaceous plants than in woody species, and in many cases no clear

18   association exists. Floral competence is the most reliable determinant that can be used to

19   distinguish between plants that are juvenile or adult. Within the context of this work, juvenility is

20   defined and measured by insensitivity to long day photoperiods, which would induce flowering

21   in adult plants.

22       The greatest advances in our understanding of the genetic regulation of plant

23   developmental transitions have derived from studying the vegetative-to-reproductive phase transition in several dicot and monocot plant species. This has led to the elucidation of multiple environmental and endogenous pathways that promote, enable and repress floral induction (Massiah 2007; Jackson 2009; Matsoukas, Massiah & Thomas 2012). The photoperiodic pathway [also known as the long day (LD) pathway] is known for its promotive effect by relaying light and photoperiodic timing signals to floral induction (reviewed in Matsoukas *et al.* 2012). This pathway involves genes such as *PHYTOCHROMES* (*PHYs*) and *CRYPTOCHROMES* (*CRYs*), which are involved in the regulation of light signal inputs. Genes such as *GIGANTEA* (*GI*), *CIRCADIAN CLOCK ASSOCIATED1* (*CCA1*) and *LATE ELONGATED HYPOCOTYL* (*LHY*) are components of the circadian clock, whereas *CONSTANS* (*CO*), *FLOWERING LOCUS T* (*FT*), *TWIN SISTER OF FT* (*TSF*) and *FLOWERING LOCUS D* (*FD)* encode proteins that specifically regulate floral induction. The action of the photoperiodic pathway ultimately converges to control the expression of so-called floral pathway integrators (FPIs), which include FT (Kardailsky *et al.* 1999; Kobayashi *et al.* 1999), TSF (Yamaguchi *et al.* 2005), SUPPRESSOR OF CONSTANS1 (SOC1; Yoo *et al.* 2005) and AGAMOUS-LIKE24 (AGL24; Lee *et al.* 2008; Liu *et al.* 2008). These act on floral meristem identity (FMI) genes *LEAFY* (*LFY*; Lee *et al.* 2008), *FRUITFUL* (*FUL*; Melzer *et al.* 2008) and *APETALA1* (*AP1*; Wigge *et al.* 2005; Yamaguchi *et al.* 2005), which result in initiation of flowering. Under short day (SD) conditions transcription of *FT*, the major output of the photoperiodic pathway, is repressed. However, as plant growth and development proceeds, *FT* expression levels show a clear increase (Yanovsky & Kay 2002).

On the other hand, pathways that enable floral induction regulate the expression of floral repressors or translocatable florigen antagonists, known as antiflorigens (Matsoukas *et al.* 2012). The pathways that regulate the floral repressor FLOWERING LOCUS C (FLC) are the best-

 

1  characterized (Michaels 2009). In addition, genetic analysis suggests that genes such as

2  *TERMINAL FLOWER1* (*TFL1*; Bradley *et al.* 1997), *LIKE HETEROCHROMATIN PROTEIN1*

3  (*LHP1, TFL2*; Gaudin *et al.* 2001; Kotake *et al.* 2003), *TEMPRANILLO* (*TEM1, TEM2*;

4  Castillejo & Pelaz 2008; Osnato *et al.* 2012) and *HASTY1* (*HST1*; Telfer & Poethig 1998) extend

5  the vegetative phase by repressing the FPIs. Functional analysis of the *hst1 Arabidopsis* mutant

6  reveals that the juvenile-to-adult phase transition is accompanied by a decrease in miR156

7  abundance and a concomitant increase in abundance of miR172, as well as the *SQUAMOSA*

8  *PROMOTER BINDING PROTEIN-LIKE* (*SPL*) transcription factors (TFs). Expression of

9  miR172 activates *FT* transcription in leaves through repression of AP2-like transcripts

10 *SCHLAFMÜTZE* (*SMZ*)*, SCHNARCHZAPFEN* (*SNZ*) and *TARGET OF EAT 1-3* (*TOE1-3*; Jung

11 *et al.* 2007; Mathieu *et al.* 2009), whereas the increase in *SPLs* at the SAM, leads to the

12 transcription of FMI genes (Wang, Czech & Weigel 2009; Yamaguchi *et al.* 2009). The FMI

13 genes trigger the expression of floral organ identity genes (Causier, Schwarz-Sommer & Davies

14 2010), which function in a combinatorial fashion to specify the distinct floral organ identities.

15     Overexpression of *CORNGRASS1*, a tandem miR156 locus, prolongs juvenility and delays

16 time to flowering in response to starch catabolism (Chuck *et al.* 2007; Gandikota *et al.* 2007).

17 Genetic and physiological approaches have demonstrated an involvement of starch with/without

18 an interaction with other plant signal transduction pathways in control of floral induction

19 (Corbesier, Lejeune & Bernier 1998; Dijken, Schluepmann & Smeekens 2004; Chuck *et al.*

20 2011; Wahl *et al.* 2013). In *Arabidopsis*, mutation in loci such as *PHOSPHOGLUCOMUTASE1*

21 (*PGM1*; Caspar, Huber & Somerville 1985; Caspar *et al.* 1991), *ADP GLUCOSE*

22 *PYROPHOSPHORYLASE1* (*ADG1*; Lin *et al.* 1988; Wang *et al.* 1998), *STARCH-EXCESS1*

23 (*SEX1*; Yu *et al.* 2001), *SEX4* (Zeeman *et al.* 1998), *CHLOROPLASTIC β-AMYLASE3* (*BAM3*;

1    Lao *et al.* 1999) and *GI* (Eimert *et al.* 1995) alter the rate of starch synthesis, accumulation or

2    mobilization conferring late flowering phenotypes under non-inductive SD conditions. The late-

3    flowering phenotype of starch-deficient mutants in SDs can be rescued by exogenous sucrose

4    application (Corbesier *et al.* 1998; Yu *et al.* 2000; Xiong *et al.* 2009). In LDs or under constant

5    light conditions, the extended daily periods partially suffice to supply enough sucrose to the

6    SAM and the starch-impaired mutants flower similar to WT.

7    *Arabidopsis* mutants allow the investigation of functional interaction between genes

8    involved in different genetic pathways, revealing the complex genetic and physiological

9    regulatory networks that orchestrate developmental transitions in plants. The objective of this

10   study was to investigate the physiological and genetic mechanisms that regulate floral

11   incompetence during juvenility in *Arabidopsis*. Using an experimental assay that allows the

12   length of the juvenile phase to be estimated based on attainment of floral competence, and

13   examination of mutants impaired in different genetic pathways, we demonstrated that multiple

14   inputs influence the timing of the juvenile-to-adult phase transition.

16   **MATERIALS AND METHODS**

18   **Plant material and growth conditions**

20   Mutant and WT *Arabidopsis* plants used were in Columbia-0 (Col-0), Landsberg *erecta*-0 (L*er*-

21   0) and Wassilewskija-4 (Ws-4) backgrounds. The background and stock number of each

22   genotype used in this study is listed in Supplementary Table S1. Mutant and WT seeds were

1    sown into Plantpak P24 module trays containing Levingtons F2 compost. Plants were grown in

2    controlled environment cabinets (Saxcil®, Chester, UK) under 100 $\mu$mol m$^2$ s$^{-1}$ PAR at 22 ±

3    0.5°C and 70 ± 2% relative humidity. When 50% of seedlings emerged, the trays with the

4    seedlings were transferred into growth cabinets (Saxcil®, Chester, UK) and the daylength

5    treatments initiated. Seven to nine replicate plants were transferred every day from SD to LD

6    conditions with the exception of plants grown in continuous SD and LD conditions where 16

7    replicate plants were used.

9    **Light sources and spectral measurements**

11    To ensure *Arabidopsis* plants under LD conditions received similar irradiance to those grown

12    under SD conditions, photoperiod was artificially increased without modifying the total quantity

13    of light available for photosynthesis, by extending the SD treatment with very low intensity

14    wavelengths that are less efficient for photosynthesis and more efficient for a photoperiodic

15    response. Short Day (SD) conditions (8 h d$^{-1}$; 100 $\mu$mol m$^2$ s$^{-1}$ PAR) were achieved using a

16    combination of fluorescent (General Electric 60W, HU) and incandescent (Philips 32W, NL)

17    light tubes. Long Day (LD) conditions (16 h d$^{-1}$ light) consisted of a combination of fluorescent

18    (General Electric 60W, HU) and incandescent (Philips 32W, NL) light for the first 8 h d$^{-1}$ (94

19    $\mu$mol m$^2$ s$^{-1}$ PAR) and low intensity (6 $\mu$mol m$^2$ s$^{-1}$ PAR) incandescent (Philips 32W, NL) light

20    for the 8 h d$^{-1}$ extension. Light quality and quantity were measured with an EPP 2000 Fiber

21    Optic Spectrometer (StellarNet Inc. USA).

**Estimation of juvenile phase length**

An analytical approach, which is based on floral competence, was used to estimate the length of the juvenile phase in *Arabidopsis* plants grown under different experimental conditions. The approach determines the phases of photoperiod sensitivity by conducting transfer experiments in which plants are transferred from SD to LD conditions at regular intervals, from seedling emergence to flowering. The approach enables the analysis of the photoperiod-insensitive juvenile vegetative phase and photoperiod-sensitive floral inductive phase of plant development. The length of these developmental phases were calculated based on the number of rosette leaves and number of days from 50% of seedling emergence at the appearance of the floral bolt at 1 cm height. Flowering time data obtained from the transfer experiments were analyzed by fitting the logistic curve, estimating the maximum slope and then fitting the lag time and stationary phase lines using the NON-LINEAR REGRESSION ANALYSIS directive of Sigma Plot 12®. The lag time and stationary phase lines were calculated by the upper and lower asymptote of the logistic curve. Analysis of variance (ANOVA) for a randomized complete block design was carried out for all data obtained using Sigma Plot 12® (Systat Software, Chicago, USA).

**Enzymatic assay of sucrose, reducing sugars and starch**

For analyses and quantification of glucose, fructose and sucrose, plant material was sampled and immediately frozen in liquid nitrogen. The freeze-dried materials were ground and 50-100 mg used for analysis. Sugars were determined enzymatically in EtOH extracts at 340 nm by a

1    UV/VIS V530 JASCO spectrophotometer, after digestions with *β*-FRUCTOSIDASE,

2    HEXOKINASE (HXK), GLUCOSE-6-PHOSPHATDEHYDROGENASE and

3    PHOSPHOGLUCOSE ISOMERASE, using the EZS 864[+] kit (Diffchamb; Lyon, FR), following

4    the manufacturer's guidelines. Starch was determined from the pellets of the soluble sugar

5    extractions after extensive washing with water. Two ml water was added per pellet, re-suspended

6    and centrifuged at 3 000 x g for 5 min. The supernatant was removed and the pellet re-extracted

7    twice using the same procedure. Starch from the air-dried pellets was quantitatively dissolved in

8    dimethyl sulfoxide (DMSO). Pellets were resuspended in 85% (v/v) DMSO and heated for 30

9    min at 90°C. After cooling, 8 M HCl was added and the solution incubated for a further 30 min

10   at 60°C. The sample was then centrifuged at 4 000 x g for 15 min. After adjusting pH to 4.5 with

11   5 M NAOH, the starch was precipitated with EtOH (96%; v/v) part of the suspension was

12   digested with AMYLOGLUCOSIDASE and HXK/GLUCOSE-6-

13   PHOSPHATDEHYDROGENASE. Starch was determined by a UV/VIS V530 JASCO (Easton,

14   USA) spectrophotometer at 340 nm, using the EnzyPlus™ determination kit (Diffchamb),

15   following the manufacturer's guidelines.

17   **RESULTS**

19   **Defining the juvenile phase length in *Arabidopsis* genotypes**

21   Transfer of plants between non-inductive SD and inductive LD photoperiods and measurement

22   of flowering times allowed the length of juvenile phase to be measured. Three *Arabidopsis* wild

1    type (WT) accessions and several mutants impaired in different genetic pathways were exploited

2    (Supplementary Table S1). Among the WT accessions under SD conditions Ws-4 showed the

3    earliest flowering phenotype, whilst Col-0 exhibited the latest flowering (Fig. 1). However,

4    under LD conditions the three accessions flowered similarly. Under SDs starch metabolism

5    mutants *pgm1, adg1-1, bam3, sex1* and *sex4* flowered significantly later than Col-0 WT (Fig. 1

6    A), whilst, under LDs only *adg1-1* and *bam3* maintained this late flowering phenotype. In

7    contrast, genotypes impaired in sugar sensing and signaling *glucose insensitive1* (*gin1*; *aba2*),

8    *gin2* (*hxk1*) and *gin6* (*abi4*) flowered with their respective WT under LDs, but earlier under SD

9    conditions (Fig. 1 B). The floral-repression pathway mutants *hst1*, *tfl1* and *lhp1* (*tfl2*) were early

10   flowering under SD conditions, whilst under LDs they flowered with their respective WT (Fig. 1

11   B, C).

12         The three *Arabidopsis* WT accessions displayed differences in juvenile phase length (Fig.

13   2; Table 1). Plants transferred from SDs to LDs whilst juvenile, flowered similarly to plants

14   receiving constant LDs. A linear increase in leaf number and days to flower with successive

15   transfer date can be seen for all the genotypes transferred after the end of juvenile phase (Fig. 2).

16   This illustrates the delay in inflorescence initiation caused by extended time spent in non-

17   inductive SD conditions. Seedlings of Ws-4 and L*er*-0 WT were insensitive to photoperiod for

18   longer periods after their emergence than Col-0, which signifies a prolonged juvenile phase. The

19   different type of mutants tested showed different durations in the length of juvenile phase. The

20   starch deficient mutants *adg1* and *pgm1* had longer juvenile phase lengths than Col-0 WT

21   (Supplementary Fig. S1 A; Table 1). The starch-excess mutants *sex1, sex4* (data not shown) and

22   *bam3* exhibited a longer juvenile phase than Col-0 WT (Supplementary Fig. S1 B; Table 1). The

23   glucose insensitive mutants *gin1* (*aba2*), *gin2* (*hxk1*) and *gin6* (*abi4*) had shorter juvenile phases,

1    compared to their respective WT (Supplementary Fig. S2; Table 1). The floral repressors *tfl1,*

2    *hst1* and *lhp1* (*tfl2*), had a shortened juvenile phase length, compared to their respective WT

3    (Supplementary Fig. S3; Table 1).

5    <Please insert Table 1 about here>

7    **Juvenile to adult phase transition and carbohydrate relationships in mutants impaired in**

8    **starch anabolism and catabolism**

10   Having established that mutants involved in starch metabolism-related events have longer

11   juvenile phase lengths compared to WT, led to the conclusion that a starch catabolism-derived

12   signal might be involved in the juvenile-to-adult phase transition. To test this hypothesis, the

13   diurnal metabolite changes in *sex1* and *adg1* mutant seedlings were determined. Col-0 WT and

14   mutant seedlings were collected under SD and LD conditions on day 9 from emergence when all

15   were in adult phase of development.

16        Starch progressively accumulated in both Col-0 WT and *sex1* mutant as the seedlings age

17   (Supplementary Fig. S4). In Col-0 WT, photosynthate assimilates were generated in excess of

18   sink demand causing elevated starch accumulation at the end of the light period (Fig. 3 A, B). By

19   the end of 16 $h^{-1}$ light period, Col-0 WT seedlings grown under SD conditions accumulated

20   greater amounts of starch, compared to plants grown in LDs. This is due to low intensity

21   incandescent light provided in the 8 $h^{-1}$ extension in the LD treatment that is less efficient for

22   photosynthesis, and more efficient for a photoperiodic response. During the dark period, reduced sucrose content triggered a degradation of starch, which was almost fully remobilized by the end of the dark period (Fig. 3 A, B). During the scotoperiod, seedlings growing in SD conditions had a slightly faster rate of starch degradation than plants growing in LD conditions. However, at the end of the dark period, similar starch contents were determined in Col-0 WT grown under both photoperiods. In contrast, *sex1* mutant showed high starch content throughout the day/night cycle and less diurnal variation under both SD and LD conditions, compared to Col-0 WT.

Determination of soluble carbohydrates extracted from *sex1* mutant seedlings showed that glucose, sucrose and fructose accumulated in large amounts during the day, relative to Col-0 WT (Fig. 3 C, D, E). With the start of 8 h$^{-1}$ light extension with low intensity incandescent light, soluble carbohydrates were depleted in WT and *sex1*. However, sucrose accumulation in *sex1* mutant seedlings were slightly reduced compared to Col-0 WT sucrose levels, at the end of dark period (Fig. 3 C).

No starch accumulated in *adg1* mutant seedlings, irrespective of daylength (Fig. 4 A, B) or developmental phase (data not presented). Compared to Col-0 WT, *adg1* mutant seedlings accumulated considerable amounts of sucrose (Fig. 4 C, D), glucose (Fig. 4 E, F) and fructose (Fig. 4 G, H) during the day, rather than being used for biosynthesis, growth and development. During the dark period, soluble carbohydrates were depleted in *adg1,* in a pattern similar to that of *sex1* (Fig. 3 C, D, E). Noticeably, at the end of dark period lesser amounts of sucrose remained in *adg1* mutant seedling grown under both photoperiods, compared to sucrose levels of Col-0 WT (Fig. 4 C, D).

**DISCUSSION**

2    Estimation of the juvenile phase length of a number of mutants acting in different genetic

3    pathways has led to a model describing a simplified integrated network of pathways that

4    quantitatively control the timing of the juvenile-to-adult phase transition. This model divides the

5    genetic pathways into those that enable the juvenile-to-adult phase transition and those that

6    promote it (Fig. 5).

8    **Defining the juvenile phase length in *Arabidopsis* WT accessions**

10    In this study, estimates on the length of juvenile phase in Col-0, L*er*-0 and Ws-4 suggested that

11    *Arabidopsis* WT accessions differ in the length of the juvenile vegetative phase. Despite the

12    hastened juvenile phase of Col-0, its late flowering phenotype in SD conditions might be

13    attributed to a prolonged photoperiod sensitive phase. Conversely, in Ws-4, despite its prolonged

14    juvenile phase compared to the other two WT accessions, the early flowering phenotype under

15    LD conditions might be due to a shortened photoperiod sensitive phase. These indicate that the

16    juvenile and adult phases of plant development can vary independently.

18    **Defining the juvenile phase length in starch deficient mutants**

20    The developmental differences between WT and starch-deficient mutants of the same

21    chronological age reveal the importance of transitory starch for normal growth and development.

22    The    *adg1*    has    no    detectable    ALPHA-D-GLUCOSE-1-PHOSPHATE    ADENYL

1 TRANSFERASE (AGP) activity, as it is deficient in the small subunit protein ADP GLUCOSE

2 PYROPHOSPHORYLASE (Lin *et al.* 1988; Wang *et al.* 1998). The *pgm1* mutant is unable to

3 synthesize starch due to inactivation of the chloroplastic isozyme of PGM (Caspar *et al.* 1985;

4 Caspar *et al.* 1991). With very low starch levels the rate of growth and net photosynthesis of both

5 mutants, and Col-0 WT are indistinguishable when the genotypes are grown under continuous

6 fluorescent light conditions (data not presented). However, under SD and LD conditions the

7 growth of *adg1* and *pgm1* is impaired and flowering is significantly delayed compared to Col-0

8 WT. It has been demonstrated that vernalization completely suppresses the late flowering

9 phenotype of *pgm1* (Bernier *et al.* 1993; Eimert *et al.* 1995) suggesting that the late-flowering

10 phenotype observed in starch deficient mutants is not due to the defect in starch accumulation

11 and slow growth rates, but more to their inability to mobilize the stored carbohydrates during the

12 scotoperiod. It has also been demonstrated that maltose and glucose are the two major forms of

13 carbon exported from chloroplasts during the scotoperiod as a result of starch catabolism

14 (Servaites & Geiger 2002; Weise, Weber & Sharkey 2004). Maltose is exported by MALTOSE

15 EXPORTER1 (MEX1; Niittyla *et al.* 2004), whereas hexokinase HXK operates as a glucose

16 sensor (Moore *et al.* 2003). The long juvenile phase length of starch-deficient mutants compared

17 to Col-0 WT provides evidence for the involvement of starch catabolism-related events in the

18 juvenile-to-adult phase transition in *Arabidopsis*.

20 **Defining the juvenile phase length in starch-excess mutants**

22 Mutants that are unable to catabolize starch provide a valuable tool to study the scotoperiodic

23 effects of carbon exported from chloroplasts on plant developmental transitions. In WT, starch is

1   degraded by phosphorylating enzymes to maltodextrin, which is then, converted to maltose and

2   glucose by BAM and DPE1 in the chloroplast for scotoperiodic export (Lao *et al.* 1999;

3   Critchley *et al.* 2001; Scheidig *et al.* 2002). Reduced activity of GLUCAN WATER DIKINASE

4   and SEX4 lead to a reduced rate of starch breakdown and in the accumulation of high levels of

5   starch in *sex1* and *sex4* mutants, respectively (Kotting *et al.* 2005; Kotting *et al.* 2009). The

6   *bam3* mutant leads also to a starch-excess phenotype (Lao *et al.* 1999). Mutants with starch-

7   excess phenotypes were late flowering compared to WT, flowering later in SD than they do in

8   LD conditions. Furthermore, *sex1, sex4* (data not shown) and *bam3* mutants displayed prolonged

9   periods of photoperiod insensitivity after seedling emergence, signifying longer juvenile phase

10  lengths than Col-0. However, the longer juvenile phase length of starch-excess mutants,

11  compared to Col-0 provides a further piece of evidence for the involvement of starch catabolism-

12  related events in the transition within the vegetative phase in *Arabidopsis*. It is plausible that

13  plants in the juvenile phase may require starch accumulation to reach a threshold level, in order

14  to sustain a steady supply of maltose and/or sucrose during the scotoperiod to undergo the

15  juvenile-to-adult phase transition. This is supported by observations in development of *mex1* and

16  *dpe-1 mex-1* double mutant. Both mutants are very small and pale and under normal growth

17  conditions, often fail to reach a mature developmental state (Niittyla *et al.* 2004). However, this

18  severe phenotype can only partially be rescued by supplying both mutants with sucrose (Stettler

19  *et al.* 2009).

21  **Defining the juvenile phase length in mutants involved in carbohydrate-hormone**

22  **interactions**

1   *Arabidopsis* mutants showing sugar insensitive phenotypes represent a valuable tool in

2   unraveling sugar-response pathways affecting developmental transitions in plants. The

3   exploitation of *gin1* (*aba2*) mutant in photoperiod transfer experiments revealed an early

4   flowering phenotype and hastened juvenile phase length, compared to L*er*-0. *GLUCOSE*

5   *INSENSITIVE1* (*ABA2*) encodes a unique short-chain DEHYDROGENASE/REDUCTASE that

6   is required for ABA synthesis. A phenotype similar to *gin1* (*aba-2*) has been determined in the

7   *gin2* (*hxk1*) mutant. *GLUCOSE INSENSITIVE2* (*HXK1*) encodes an HXK that functions as a

8   glucose sensor to integrate nutrient, light intensity, and hormone-signaling systems for

9   controlling plant development in response to environmental conditions (Moore *et al.* 2003). The

10  *gin6* (*abi4*) mutant in the Col-0 background contains a T-DNA insertion in the promoter of the

11  At2g40220 locus, which encodes an APETALA-2 domain TF (Finkelstein *et al.* 1998; Arenas-

12  Huertero *et al.* 2000). The short juvenile phase length of *gin6* and other glucose insensitive

13  mutants demonstrates the involvement of these loci in the juvenile-to-adult phase transition in

14  *Arabidopsis*. Noticeably, it has been shown that in addition to *GIN6*, several mutants insensitive

15  to sucrose are allelic to *abi4* (Huijser *et al.* 2000; Rook *et al.* 2001). This might disclose the tight

16  interplay between sugar and ABA phytohormone pathway in the regulation of the juvenile-to-

17  adult phase transition through multiple pathways (Choi *et al.* 2000; Domagalska *et al.* 2010).

19  **Defining the juvenile phase length in mutants acting as floral repressors**

21  Floral incompetence during the juvenile phase has led to the hypothesis that the underlying

22  mechanism of juvenility may involve activities of strong floral repressors based at the leaf or at

23  SAM. *Terminal flower1* (*tfl1*), *lhp1* (*tfl2*) and *hst1* mutants were shown to flower early under non-inductive SD conditions compared to their respective WT. Assessment of juvenility in *tfl1*, *hst1* and *lhp1* (*tfl2*) seedlings showed they had shortened juvenile phase lengths compared to their respective WT accessions, which is significantly longer. This provides evidence for the involvement of these three loci in the juvenile-to-adult-phase transition in *Arabidopsis*. Genetic and molecular approaches have identified the functions of *TFL1*, *HST1* and *LHP1* (*TFL2*). Transcripts of these genes are detected in all plant tissues (Schmid *et al.* 2005). However, it has been shown that they mainly act at the SAM by regulating the FMI genes.

*HASTY1* is the *Arabidopsis* ortholog of the miRNA nuclear export receptor EXPORTIN5 (Bollman *et al.* 2003). *HASTY1* (*HST1*) was isolated in a screen for mutations that accelerated the vegetative phase change with respect to leaf morphological traits. Functional analysis of *hst* mutations revealed a role for *miR156* and *miR172* in synchronization and harmonization of the juvenile-to-adult phase transition (Wu & Poethig 2006; Chuck *et al.* 2007; Jung *et al.* 2007; Mathieu *et al.* 2007). It has been shown that the levels of *miR156* and *miR172* exhibit contrasting expression patterns (Chuck *et al.* 2007; Wu *et al.* 2009). As age proceeds, the decline of *miR156* levels, and the increase in levels of *miR172* and certain *SPL* genes, leads to the activation of FT in leaves, whereas the increase in *SPLs* in the meristem leads to the activation of FPIs and FMI genes that promote the transition to flowering (Wang *et al.* 2009; Yamaguchi *et al.* 2009).

*TERMINAL FLOWER1* (*TFL1*) has been demonstrated to function as a signal to coordinate shoot meristem identity by regulating the FMI genes *LFY* and *AP1* (Ratcliffe, Bradley & Coen 1999; Conti & Bradley 2007). Further support for the involvement of *TFL1* in the vegetative phase transition is derived from the study of Bradley *et al.* (1997) on inflorescence commitment in *Arabidopsis*. By applying LD to SD transfer of seedlings, the juvenile phase length of *tfl1* was shortened compared to L*er*-0 WT.

1    LIKE HETEROCHROMATIN PROTEIN1 (LHP1/TFL2), functions as a negative

2    regulator during the vegetative-to-reproductive phase transition by repressing the expression of

3    *FT*, but with no effect on expression of the other FPIs (Gaudin *et al.* 2001; Kotake *et al.* 2003). It

4    has been demonstrated that HP1, with which LHP1 shares homology, maintains genes in a

5    transcriptionally inactive state by remodeling chromatin structure in the heterochromatin region

6    (Nakahigashi *et al.* 2005). It is plausible that not only the activators but also the repressors are

7    required for the precise synchronization and harmonization of the juvenile-to-adult phase

8    transition in *Arabidopsis*. Taken together, these results suggest that regulation of juvenility might

9    be through repression of FPIs transcription by antiflorigenic molecules such as TFL1, HST1 and

10   LHP1 (TFL2).

12   **Juvenile-to-adult phase transition and carbohydrate relationships**

14   Starch catabolism-related events might be the cause of the prolonged juvenile phase in starch-

15   deficient and starch-excess mutants. It has been proposed that the inhibition of growth in starch-

16   deficient mutants is primarily caused by a disturbance of metabolism and growth, which is

17   triggered by a transient period of sugar depletion during the scotoperiod (Bernier *et al.* 1993).

18   Furthermore, it has been demonstrated that the expression of hundreds of genes is altered in the

19   *pgm1* starch-deficient mutant at the end of the dark period, compared to WT at the same time

20   (Thimm *et al.* 2004). This includes many genes that are required for nutrient assimilation,

21   biosynthesis and growth. Taken together, these results suggest that soluble carbohydrate

22   depletion during the dark period leads to marked changes in gene expression stimulating an

23   inhibition of carbohydrate utilization, which could directly affect the juvenile-to-adult phase

1    transition. However, in addition to transitory starch providing a source of carbon for growth

2    during the following night (Thimm *et al.* 2004) and for the beginning of the next light period, it

3    may also act as an over-flow for newly assimilated carbon (Stitt & Quick 1989), when

4    assimilation exceeds the demand for sucrose. This mechanism is inactivated in the *adg1* mutant,

5    as demonstrated by the elevated soluble carbohydrate levels, and hardly any starch at the end of

6    the light period, in *adg1* mutant seedlings.

7        The *sex1* mutant is known for its impaired ability to catabolize starch (Caspar *et al.* 1991;

8    Yu *et al.* 2001). The lack of starch turnover in *sex1* has an influence on general carbohydrate

9    availability, reducing the amount of sucrose and maltose contents (Chia *et al.* 2004; Niittyla *et*

10    *al.* 2004) at the end of scotoperiod. The importance of temporal availability of maltose in the

11    regulation of plant growth has already been demonstrated (Niittyla *et al.* 2004; Stettler *et al.*

12    2009). Furthermore, despite the fact that *sex1* and *adg1* mutants being impaired in different

13    genetic pathways their metabolism and growth inhibition might be triggered by a transient period

14    of soluble carbohydrate depletion during the scotoperiod. It is possible that both mutants

15    function in the same physiological pathway controlling juvenility. As with *adg1*, it is possible

16    that in *sex1* carbohydrate depletion during the dark leads to critical changes in gene expression

17    stimulating an inhibition of carbohydrate utilization with direct effects on the length of the

18    juvenile phase.

19        A number of physiological, biochemical and molecular approaches have shown that early

20    growth and development in *Arabidopsis* seedlings can be arrested in the presence of high glucose

21    and sucrose levels. Several developmental characteristics are subject to high-level soluble

22    carbohydrate repression (Jang *et al.* 1997; Arenas-Huertero *et al.* 2000; Gibson 2000; Gazzarrini

23    & McCourt 2001). The physiological rationale for soluble carbohydrate repression during the early phases of plant development could be that elevated soluble carbohydrate accumulation levels reflect suboptimal growth conditions (Lopez-Molina, Mongrand & Chua 2001). Therefore, inhibition of developmental programs such as the juvenile-to-adult phase transition in starch-deficient and starch-excess mutants may result from the activation of soluble carbohydrate repression events. Based on this repression of growth response, a series of *gin* mutants have been isolated (Zhou *et al.* 1998; Moore *et al.* 2003). The finding that the *gin* mutants have a shortened juvenile phase and an early flowering phenotype further supports this hypothesis.

The data presented shows that a variety of signals act to promote and enable the juvenile-to-adult phase transition that involves both floral activators and repressors. Starch metabolism is involved in the juvenile-to-adult phase transition. Carbohydrates might be involved through their function as nutrients, osmotic regulators and signalling molecules, and/ or by their interaction with phytohormonal networks.

**FUNDING**

This work was supported by the Hellenic State Scholarships Foundation (IKY) and UK Department for Environment, Food and Rural Affairs (DEFRA).

**ACKNOWLEDGEMENTS**

We would like to thank Prof. Samuel Zeeman (ETH, Zurich) for providing the *sex4* and *bam3* mutants, Prof. Jen Sheen (Harvard Medical School) for providing the *gin* mutants and Prof.

1    Valérie Gaudin (INRA, Versailles) for providing the *lhp1* mutant.

**REFERENCES**

Arenas-Huertero F., Arroyo A., Zhou L., Sheen J. & Leon P. (2000) Analysis of Arabidopsis glucose insensitive mutants, gin5 and gin6, reveals a central role of the plant hormone ABA in the regulation of plant vegetative development by sugar. *Genes and Development*, **14**, 2085-2096.

Bernier G., Havelange A., Houssa C., Petitjean A. & Lejeune P. (1993) Physiological signals that induce flowering. *The Plant Cell*, **5**, 1147-1155.

Bollman K.M., Aukerman M.J., Park M.Y., Hunter C., Berardini T.Z. & Poethig R.S. (2003) HASTY, the Arabidopsis ortholog of exportin 5/MSN5, regulates phase change and morphogenesis. *Development*, **130**, 1493-1504.

Bradley D., Ratcliffe O., Vincent C., Carpenter R. & Coen E. (1997) Inflorescence commitment and architecture in Arabidopsis. *Science*, **275**, 80-83.

Caspar T., Huber S.C. & Somerville C. (1985) Alterations in growth, photosynthesis, and respiration in a starchless mutant of Arabidopsis thaliana (L.) deficient in chloroplast phosphoglucomutase activity. *Plant Physiology*, **79**, 11-17.

Caspar T., Lin T.P., Kakefuda G., Benbow L., Preiss J. & Somerville C. (1991) Mutants of Arabidopsis with altered regulation of starch degradation. *Plant Physiology*, **95**, 1181-1188.

Castillejo C. & Pelaz S. (2008) The balance between CONSTANS and TEMPRANILLO activities determines FT expression to trigger flowering. *Current Biology*, **18**, 1338-1343.

1   Causier B., Schwarz-Sommer Z. & Davies B. (2010) Floral organ identity: 20 years of ABCs.
2       *Seminars in Cell & Developmental Biology*, **21**, 73-79.

3   Chia T., Thorneycroft D., Chapple A., Messerli G., Chen J., Zeeman S.C., Smith S.M. & Smith
4       A.M. (2004) A cytosolic glucosyltransferase is required for conversion of starch to
5       sucrose in Arabidopsis leaves at night. *The Plant Journal*, **37**, 853-863.

6   Choi H.I., Hong J.H., Ha J.O., Kang J.Y. & Kim S.Y. (2000) ABFs, a family of ABA-responsive
7       element binding factors. *Journal of Biological Chemistry*, **275**, 1723-1730.

8   Chuck G.S., Cigan A.M., Saeteurn K. & Hake S. (2007) The heterochronic maize mutant
9       Corngrass1 results from overexpression of a tandem microRNA. *Nature Genetics*, **39**,
10      544-549.

11  Chuck G.S., Tobias C., Sun L., Kraemer F., Li C.L., Dibble D., Arora R., Bragg J.N., Vogel J.P.,
12      Singh S., Simmons B.A., Pauly M. & Hake S. (2011) Overexpression of the maize
13      Corngrass1 microRNA prevents flowering, improves digestibility, and increases starch
14      content of switchgrass. *Proceedings of the National Academy of Sciences of the United*
15      *States of America*, **108**, 17550-17555.

16  Conti L. & Bradley D. (2007) TERMINAL FLOWER1 is a mobile signal controlling
17      Arabidopsis architecture. *The Plant Cell*, **19**, 767-778.

18  Corbesier L., Lejeune P. & Bernier G. (1998) The role of carbohydrates in the induction of
19      flowering in Arabidopsis thaliana: comparison between the wild type and a starchless
20      mutant. *Planta*, **206**, 131-137.

1  Critchley J.H., Zeeman S.C., Takaha T., Smith A.M. & Smith S.M. (2001) A critical role for
2      disproportionating enzyme in starch breakdown is revealed by a knock-out mutation in
3      Arabidopsis. *The Plant Journal*, **26**, 89-100.

4  Dijken A., Schluepmann H. & Smeekens S. (2004) Arabidopsis trehalose-6-phosphate synthase
5      1 is essential for normal vegetative growth and transition to flowering. *Plant Physiology*,
6      **135**, 969-977.

7  Domagalska M.A., Sarnowska E., Nagy F. & Davis S.J. (2010) Genetic analyses of interactions
8      among gibberellin, abscisic acid, and brassinosteroids in the control of flowering time in
9      Arabidopsis thaliana. *PLoS ONE*, **5**, e14012.

10 Eimert K., Wang S.M., Lue W.I. & Chen J. (1995) Monogenic recessive mutations causing both
11      late floral initiation and excess starch accumulation in Arabidopsis. *The Plant Cell*, **7**,
12      1703-1712.

13 Finkelstein R.R., Wang M.L., Lynch T.J., Rao S. & Goodman H.M. (1998) The Arabidopsis
14      abscisic acid response locus ABI4 encodes an APETALA 2 domain protein. *The Plant
15      Cell*, **10**, 1043-1054.

16 Gandikota M., Birkenbihl R.P., Hohmann S., Cardon G.H., Saedler H. & Huijser P. (2007) The
17      miRNA156/157 recognition element in the 3' UTR of the Arabidopsis SBP box gene
18      SPL3 prevents early flowering by translational inhibition in seedlings. *The Plant Journal*,
19      **49**, 683-693.

20 Gaudin V., Libault M., Pouteau S., Juul T., Zhao G., Lefebvre D. & Grandjean O. (2001)
21      Mutations in LIKE HETEROCHROMATIN PROTEIN1 affect flowering time and plant
22      architecture in Arabidopsis. *Development*, **128**, 4847-4858.

1   Gazzarrini S. & McCourt P. (2001) Genetic interactions between ABA, ethylene and sugar
2         signaling pathways. *Current Opinion in Plant Biology*, **4**, 387-391.

3   Gibson S.I. (2000) Plant sugar-response pathways. Part of a complex regulatory web. *Plant
4         Physiology*, **124**, 1532-1539.

5   Huijser C., Kortstee A., Pego J., Weisbeek P., Wisman E. & Smeekens S. (2000) The
6         Arabidopsis SUCROSE UNCOUPLED-6 gene is identical to ABSCISIC ACID
7         INSENSITIVE-4: involvement of abscisic acid in sugar responses. *The Plant Journal*,
8         **23**, 577-585.

9   Jackson S. (2009) Plant responses to photoperiod. *New Phytologist*, **181**, 517-531.

10  Jang J.C., Leon P., Zhou L. & Sheen J. (1997) Hexokinase as a sugar sensor in higher plants. *The
11        Plant Cell*, **9**, 5-19.

12  Jung J.H., Seo Y.H., Seo P.J., Reyes J.L., Yun J., Chua N.H. & Park C.M. (2007) The
13        GIGANTEA-regulated microRNA172 mediates photoperiodic flowering independent of
14        CONSTANS in Arabidopsis. *The Plant Cell*, **19**, 2736-2748.

15  Kardailsky I., Shukla V.K., Ahn J.H., Dagenais N., Christensen S.K., Nguyen J.T., Chory J.,
16        Harrison M.J. & Weigel D. (1999) Activation tagging of the floral inducer FT. *Science*,
17        **286**, 1962-1965.

18  Kobayashi Y., Kaya H., Goto K., Iwabuchi M. & Araki T. (1999) A pair of related genes with
19        antagonistic roles in mediating flowering signals. *Science*, **286**, 1960-1962.

20  Kotake T., Takada S., Nakahigashi K., Ohto M. & Goto K. (2003) Arabidopsis TERMINAL
21        FLOWER2 gene encodes a heterochromatin protein1 homolog and represses both FLOWERING LOCUS T to regulate flowering time and several floral homeotic genes. *Plant and Cell Physiology*, **44**, 555-564.

Kotting O., Pusch K., Tiessen A., Geigenberger P., Steup M. & Ritte G. (2005) Identification of a novel enzyme required for starch metabolism in Arabidopsis leaves. The phosphoglucan, water dikinase. *Plant Physiology*, **137**, 242-252.

Kotting O., Santelia D., Edner C., Eicke S., Marthaler T., Gentry M.S., Comparot-Moss S., Chen J., Smith A.M., Steup M., Ritte G. & Zeeman S.C. (2009) STARCH-EXCESS4 Is a laforin-like phosphoglucan phosphatase required for starch degradation in Arabidopsis thaliana. *The Plant Cell*, **21**, 334-346.

Lao N.T., Schoneveld O., Mould R.M., Hibberd J.M., Gray J.C. & Kavanagh T.A. (1999) An Arabidopsis gene encoding a chloroplast-targeted beta-amylase. *The Plant Journal*, **20**, 519-527.

Lee J., Oh M., Park H. & Lee I. (2008) SOC1 translocated to the nucleus by interaction with AGL24 directly regulates LEAFY. *The Plant Journal*, **55**, 832-843.

Lin T.P., Caspar T., Somerville C.R. & Preiss J. (1988) A starch deficient mutant of Arabidopsis thaliana with low ADPglucose pyrophosphorylase activity lacks one of the two subunits of the enzyme. *Plant Physiology*, **88**, 1175-1181.

Liu C., Chen H., Er H.L., Soo H.M., Kumar P.P., Han J.H., Liou Y.C. & Yu H. (2008) Direct interaction of AGL24 and SOC1 integrates flowering signals in Arabidopsis. *Development*, **135**, 1481-1491.

Lopez-Molina L., Mongrand S. & Chua N.H. (2001) A postgermination developmental arrest checkpoint is mediated by abscisic acid and requires the AB15 transcription factor in Arabidopsis. *Proceedings of the National Academy of Sciences of the United States of America*, **98**, 4782-4787.

Massiah A. (2007) Understanding flowering time. *CAB Reviews: Perspectives in Agriculture, Veterinary Science, Nutrition and Natural Resources*, **2**, 1-21.

Mathieu J., Warthmann N., Kuttner F. & Schmid M. (2007) Export of FT protein from phloem companion cells is sufficient for floral induction in Arabidopsis. *Current Biology*, **17**, 1055-1060.

Mathieu J., Yant L.J., Murdter F., Kuttner F. & Schmid M. (2009) Repression of flowering by the miR172 target SMZ. *PLoS Biology*, **7**, e1000148.

Matsoukas I.G., Massiah A.J. & Thomas B. (2012) Florigenic and antiflorigenic signalling in plants. *Plant and Cell Physiology*, **53**, 1827-1842.

Melzer S., Lens F., Gennen J., Vanneste S., Rohde A. & Beeckman T. (2008) Flowering-time genes modulate meristem determinacy and growth form in Arabidopsis thaliana. *Nature Genetics*, **40**, 1489-1492.

Michaels S.D. (2009) Flowering time regulation produces much fruit. *Current Opinion in Plant Biology*, **12**, 75-80.

Moore B.D., Zhou L., Rolland F., Hall Q., Cheng W.H., Liu Y.X., Hwang I., Jones T. & Sheen J. (2003) Role of the Arabidopsis glucose sensor HXK1 in nutrient, light, and hormonal signaling. *Science*, **300**, 332-336.

Nakahigashi K., Jasencakova Z., Schubert I. & Goto K. (2005) The Arabidopsis HETEROCHROMATIN PROTEIN1 homolog (TERMINAL FLOWER2) silences genes within the euchromatic region but not genes positioned in heterochromatin. *Plant and Cell Physiology*, **46**, 1747-1756.

Niittyla T., Messerli G., Trevisan M., Chen J., Smith A.M. & Zeeman S.C. (2004) A previously unknown maltose transporter essential for starch degradation in leaves. *Science*, **303**, 87-89.

Osnato M., Castillejo C., Matias-Hernandez L. & Pelaz S. (2012) TEMPRANILLO genes link photoperiod and gibberellin pathways to control flowering in Arabidopsis. *Nature Communications*, **3**, 808.

Poethig R.S. (1990) Phase change and the regulation of shoot morphogenesis in plants. *Science*, **250**, 923-930.

Ratcliffe O.J., Bradley D.J. & Coen E.S. (1999) Separation of shoot and floral identity in Arabidopsis. *Development*, **126**, 1109-1120.

Rook F., Corke F., Card R., Munz G., Smith C. & Bevan M.W. (2001) Impaired sucrose-induction mutants reveal the modulation of sugar-induced starch biosynthetic gene expression by abscisic acid signalling. *The Plant Journal*, **26**, 421-433.

Scheidig A., Frohlich A., Schulze S., Lloyd J.R. & Kossmann J. (2002) Downregulation of a chloroplast-targeted beta-amylase leads to a starch-excess phenotype in leaves. *The Plant Journal*, **30**, 581-591.

Schmid M., Davison T.S., Henz S.R., Pape U.J., Demar M., Vingron M., Scholkopf B., Weigel D. & Lohmann J.U. (2005) A gene expression map of Arabidopsis thaliana development. *Nature Genetics*, **37**, 501-506.

1   Servaites J.C. & Geiger D.R. (2002) Kinetic characteristics of chloroplast glucose transport.
2       *Journal of Experimental Botany*, **53**, 1581-1591.

3   Stettler M., Eicke S., Mettler T., Messerli G., Hortensteiner S. & Zeeman S.C. (2009) Blocking
4       the metabolism of starch breakdown products in Arabidopsis leaves triggers chloroplast
5       degradation. *Molecular Plant*, **2**, 1233-1246.

6   Stitt M. & Quick P. (1989) Photosynthetic carbon partitioning - Its regulation and possibilities
7       for manipulation. *Physiologia Plantarum*, **77**, 633-641.

8   Telfer A. & Poethig R.S. (1998) HASTY: a gene that regulates the timing of shoot maturation in
9       Arabidopsis thaliana. *Development*, **125**, 1889-1898.

10  Thimm O., Blasing O., Gibon Y., Nagel A., Meyer S., Kruger P., Selbig J., Muller L.A., Rhee
11      S.Y. & Stitt M. (2004) MAPMAN: a user-driven tool to display genomics data sets onto
12      diagrams of metabolic pathways and other biological processes. *The Plant Journal*, **37**,
13      914-939.

14  Thomas B. & Vince-Prue D. (1984) Juvenility, photoperiodism and vernalization. In: *Advanced
15      Plant Physiology* (ed M.B. Wilkins), pp. 408–439. Pitman, London, UK.

16  Wahl V., Ponnu J., Schlereth A., Arrivault S., Langenecker T., Franke A., Feil R., Lunn J. E.,
17      Stitt M. & Schmid M. (2013) Regulation of flowering by trehalose-6-phosphate signaling
18      in Arabidopsis thaliana. *Science*, **339**, 704-707.

19  Wang J.W., Czech B. & Weigel D. (2009) miR156-regulated SPL transcription factors define an
20      endogenous flowering pathway in Arabidopsis thaliana. *Cell*, **138**, 738-749.

21  Wang S.M., Lue W.L., Yu T.S., Long J.H., Wang C.N., Eimert K. & Chen J. (1998)
22      Characterization of ADG1, an Arabidopsis locus encoding for ADPG pyrophosphorylase

small subunit, demonstrates that the presence of the small subunit is required for large subunit stability. *The Plant Journal*, **13**, 63-70.

Weise S.E., Weber A.P.M. & Sharkey T.D. (2004) Maltose is the major form of carbon exported from the chloroplast at night. *Planta*, **218**, 474-482.

Wigge P.A., Kim M.C., Jaeger K.E., Busch W., Schmid M., Lohmann J.U. & Weigel D. (2005) Integration of spatial and temporal information during floral induction in Arabidopsis. *Science*, **309**, 1056-1059.

Wu G., Park M.Y., Conway S.R., Wang J.W., Weigel D. & Poethig R.S. (2009) The sequential action of miR156 and miR172 regulates developmental timing in Arabidopsis. *Cell*, **138**, 750-759.

Wu G. & Poethig R.S. (2006) Temporal regulation of shoot development in Arabidopsis thaliana by miR156 and its target SPL3. *Development*, **133**, 3539-3547.

Xiong Y.Q., DeFraia C., Williams D., Zhang X.D. & Mou Z.L. (2009) Deficiency in a cytosolic ribose-5-phosphate isomerase causes chloroplast dysfunction, late flowering and premature cell death in Arabidopsis. *Physiologia Plantarum*, **137**, 249-263.

Yamaguchi A., Kobayashi Y., Goto K., Abe M. & Araki T. (2005) Twin Sister of FT (TSF) acts as a floral pathway integrator redundantly with FT. *Plant and Cell Physiology*, **46**, 1175-1189.

Yamaguchi A., Wu M.F., Yang L., Wu G., Poethig R.S. & Wagner D. (2009) The microRNA-regulated SBP-Box transcription factor SPL3 is a direct upstream activator of LEAFY, FRUITFULL, and APETALA1. *Developmental Cell*, **17**, 268-278.

Yanovsky M.J. & Kay S.A. (2002) Molecular basis of seasonal time measurement in Arabidopsis. *Nature*, **419**, 308-312.

Yoo S.K., Chung K.S., Kim J., Lee J.H., Hong S.M., Yoo S.J., Yoo S.Y., Lee J.S. & Ahn J.H. (2005) CONSTANS activates SUPPRESSOR OFOVEREXPRESSION OFCONSTANS 1 through FLOWERING LOCUS T to promote flowering in Arabidopsis. *Plant Physiology*, **139**, 770-778.

Yu T.S., Kofler H., Hausler R.E., Hille D., Flugge U.I., Zeeman S.C., Smith A.M., Kossmann J., Lloyd J., Ritte G., Steup M., Lue W.L., Chen J.C. & Weber A. (2001) The Arabidopsis sex1 mutant is defective in the R1 protein, a general regulator of starch degradation in plants, and not in the chloroplast hexose transporter. *The Plant Cell*, **13**, 1907-1918.

Yu T.S., Lue W.L., Wang S.M. & Chen J.C. (2000) Mutation of arabidopsis plastid phosphoglucose isomerase affects leaf starch synthesis and floral initiation. *Plant Physiology*, **123**, 319-325.

Zeeman S.C., Northrop F., Smith A.M. & ap Rees T. (1998) A starch-accumulating mutant of Arabidopsis thaliana deficient in a chloroplastic starch-hydrolysing enzyme. *The Plant Journal*, **15**, 357-365.

Zhou L., Jang J.C., Jones T.L. & Sheen J. (1998) Glucose and ethylene signal transduction crosstalk revealed by an Arabidopsis glucose-insensitive mutant. *Proceedings of the National Academy of Sciences of the United States of America*, **95**, 10294-10299.

1    **FIGURE AND TABLE LEGENDS**

3    **Figure 1.** Flowering time profiles of *Arabidopsis* WTs and mutant genotypes grown under SD

4    and LD conditions.

5    (A) Number of rosette leaves at flowering of Col-0 WT and mutants in the Col-0 background.

6    (B) Number of rosette leaves at flowering of L*er*-0 WT and mutants in the L*er*-0 background.

7    (C) Number of rosette leaves at flowering of Ws-4 WT and *lhp1-1* mutant. Dark bars denote

8    number of rosette leaves in LDs and light bars number of rosette leaves in SDs. Short day (SD)

9    conditions (8 h $d^{-1}$) were achieved using a combination of fluorescent and incandescent light.

10   Long day (LD) conditions (16 h $d^{-1}$) consisted of a combination of fluorescent and incandescent

11   light for the first 8 h $d^{-1}$ and low intensity incandescent light for the 8 h $d^{-1}$ extension. Data are

12   represented by analysis of 14 plants for each photoperiod treatment. Flowering time is expressed

13   as the number of rosette leaves from 50% of seedling emergence to the appearance of the floral

14   bolt at 1 cm height. Error bars indicate ± standard error of mean (SEM). Values followed by the

15   same letter are not significantly different according to the Student's t test at a 0.05 level of

16   significance.

1    **Figure 2.** Estimation of the juvenile phase length of Col-0 (●), L*er*-0 (○) and Ws-4 (▼)

2    accessions of *Arabidopsis* based on attainment of floral competence.

3    Plants were transferred from SDs to LDs at time intervals shown in x-axes. Points and vertical

4    error bars denote the mean and SEM of leaves at flowering of replicate plants transferred on each

5    occasion. The dashed line delimits the length of juvenile phase. Horizontal error bars denote the

6    SEM of the estimated juvenile phase length.

1    **Figure 3.** Diurnal metabolite changes in Col-0 WT and *sex1* mutant genotypes of *Arabidopsis*.

2    Diurnal changes of starch (A, B), sucrose (C), glucose (D) and fructose (E) in Col-0 and *sex1*

3    mutant genotypes. Col-0 (●) and *sex1* (○) seedlings were collected under SDs (A) and LDs (B,

4    C, D, E) on day 9 from emergence, at time intervals shown in x-axes. All results are the averages

5    of three biological replicates ± SEM. White and black bars on the top are subjective day and

6    night, while the grey bar in the LD treatment indicates the photoperiod extension with low

7    intensity incandescent light.

1    **Figure 4.** Diurnal metabolite changes in Col-0 WT and *adg1-1* mutant genotypes of *Arabidopsis*.

2    Diurnal changes of starch (A, B), sucrose (C, D), glucose (E, F) and fructose (G, E) in Col-0 and

3    *adg1-1* genotypes. Col-0 (●), and *adg1-1* (○) seedlings were collected under SDs (A, C, E, G)

4    and LDs (B, D, F, H) at time intervals shown in x-axes. All results are the averages of three

5    biological replicates ± SEM. White and black bars on the top are subjective day and night,

6    whereas the grey bar in the LD treatment indicates the photoperiod extension with low intensity

7    incandescent light.

1  **Figure 5.** A model describing a simplified-integrated network of pathways that quantitatively

2  control the timing of the juvenile-to-adult phase transition in *Arabidopsis*.

3  Mutations in different genetic pathways are grouped into those that promote (↓) and those that

4  repress (⊥) the juvenile-to-adult phase transition. The enabling pathways regulate the ability of

5  the leaf and meristem to respond to floral promotive signals from different environmental and

6  endogenous cues.

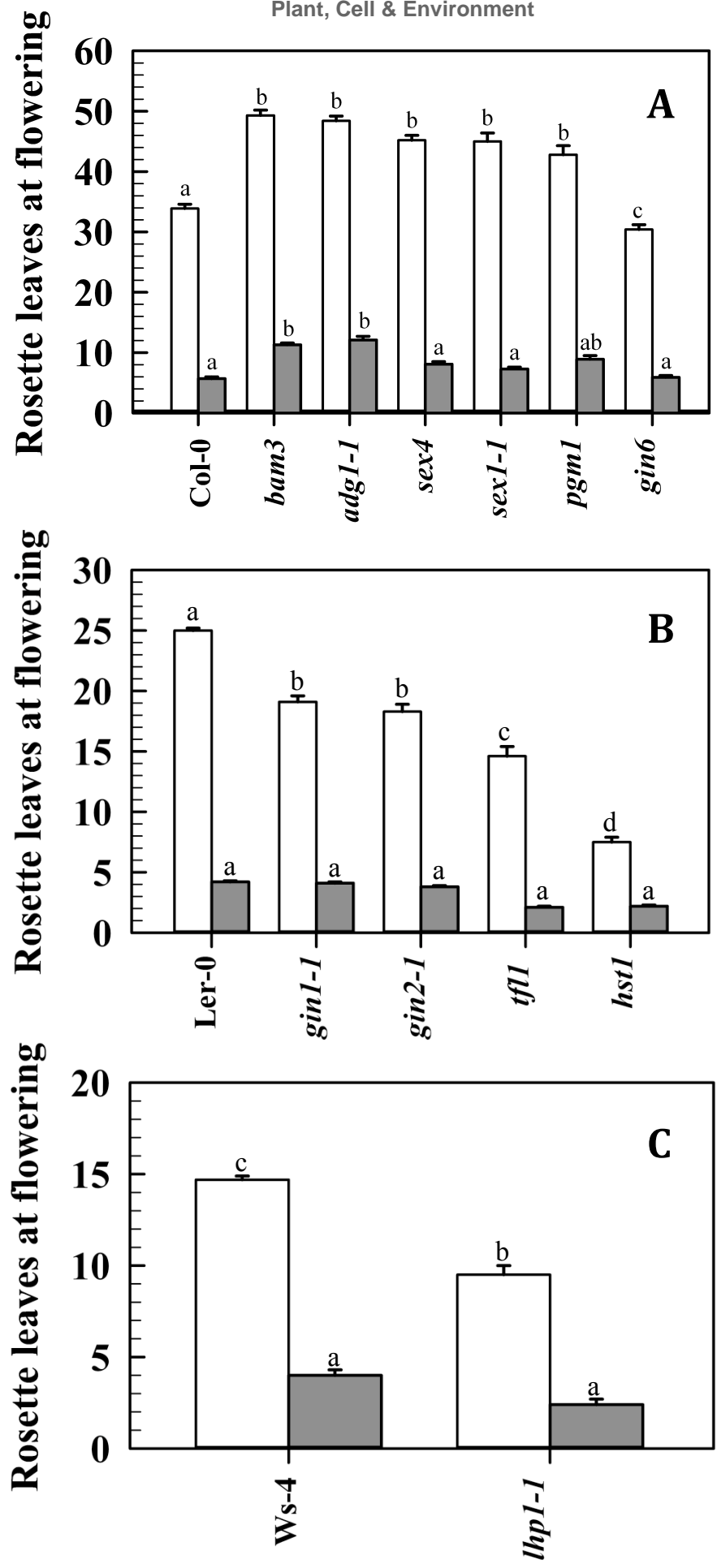

**Figure 1**

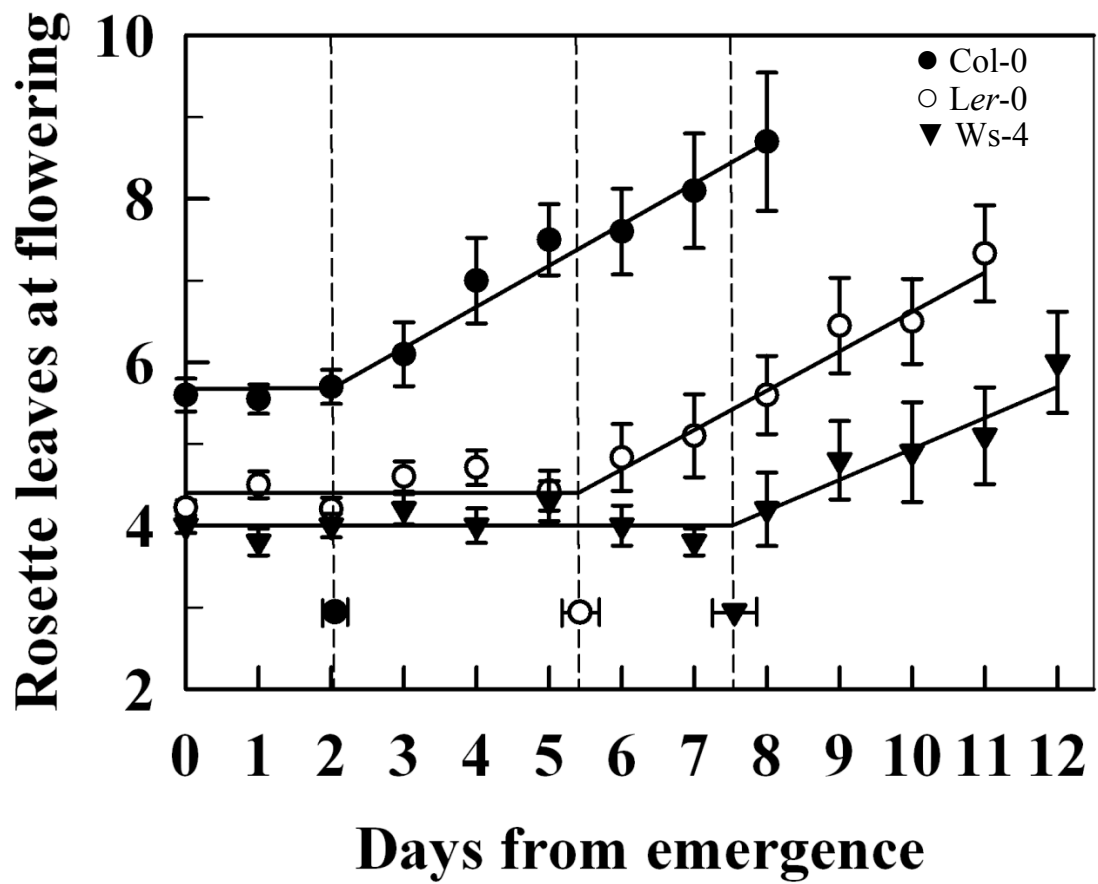

**Figure 2**

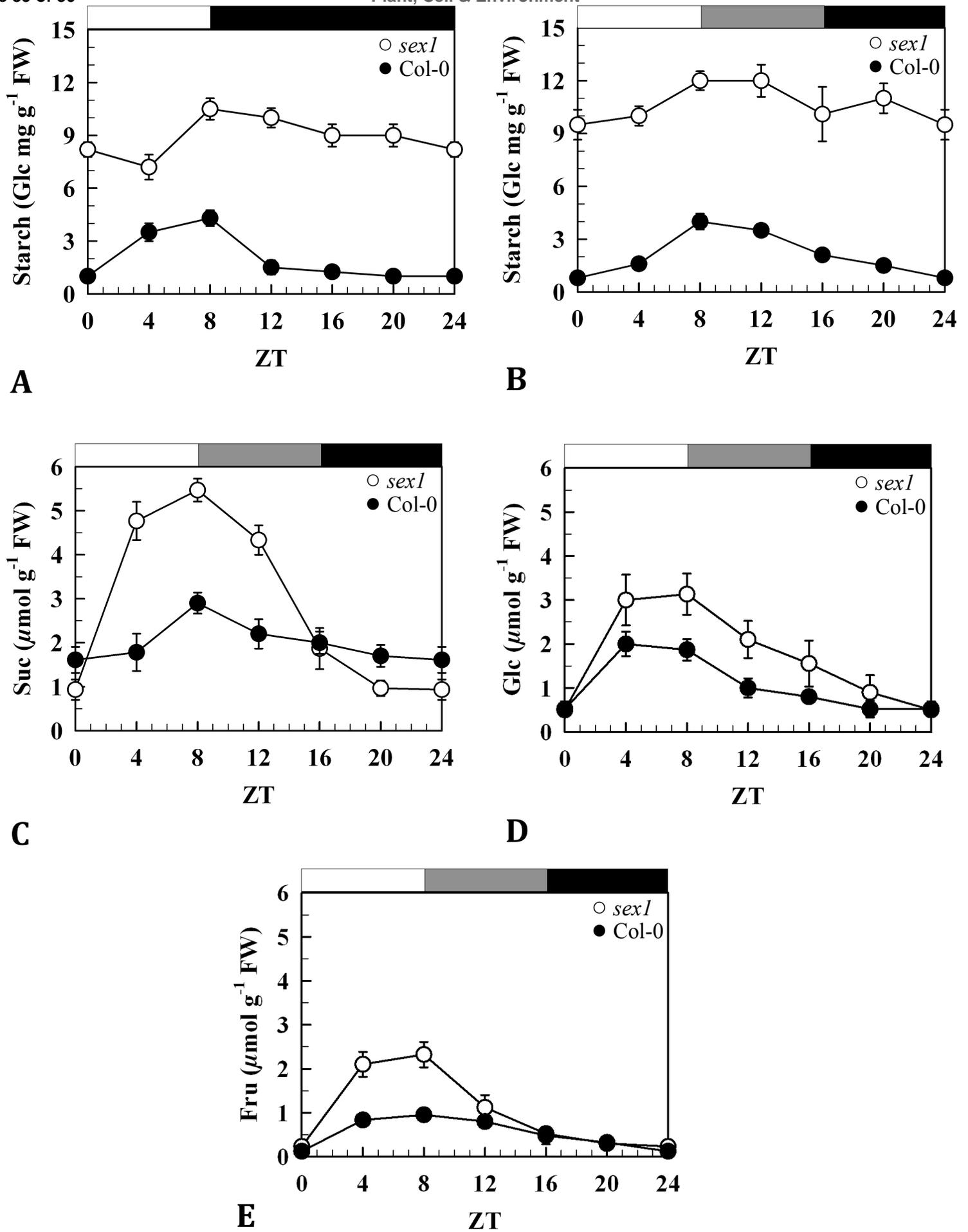

Figure 3

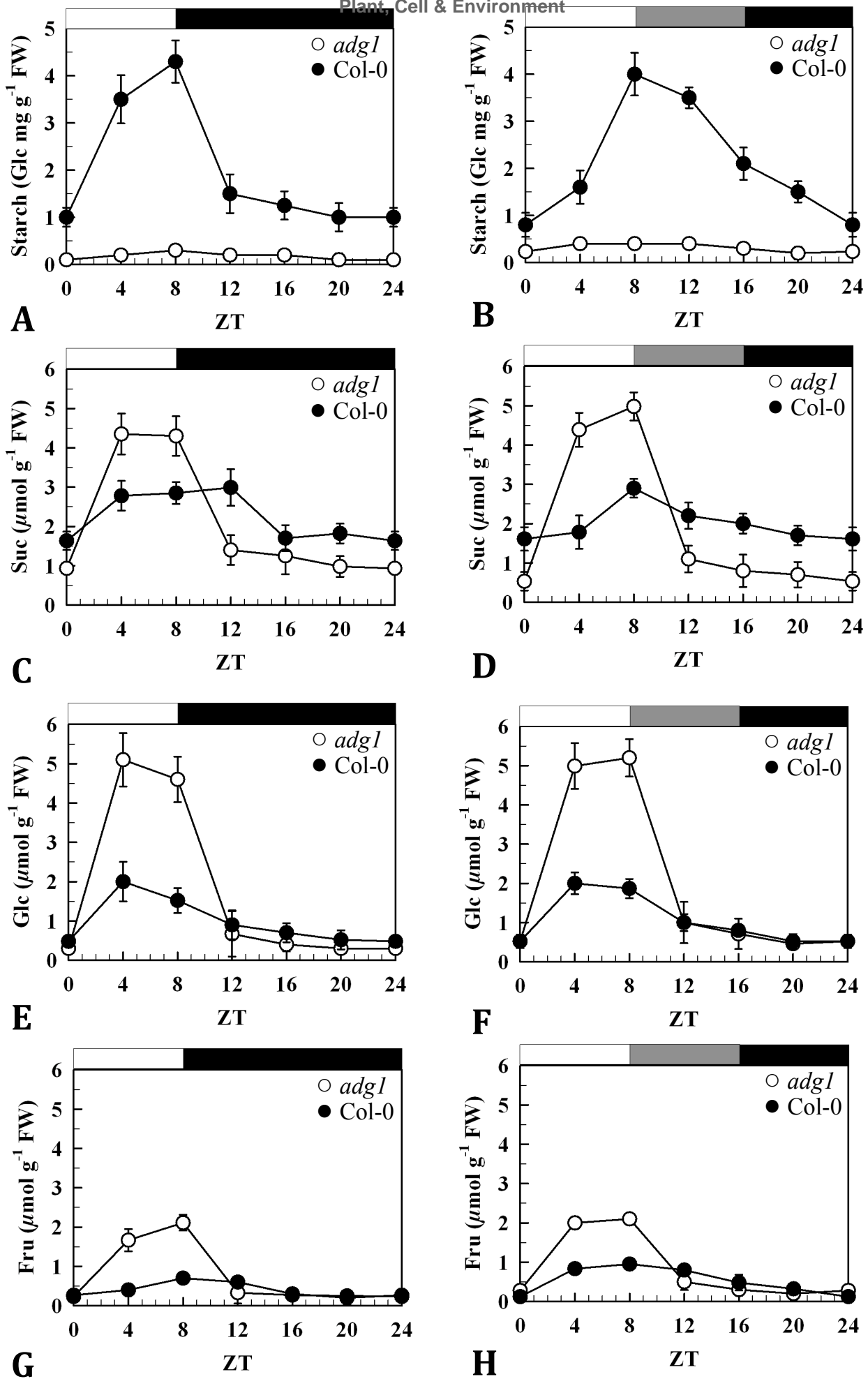

**Figure 4**

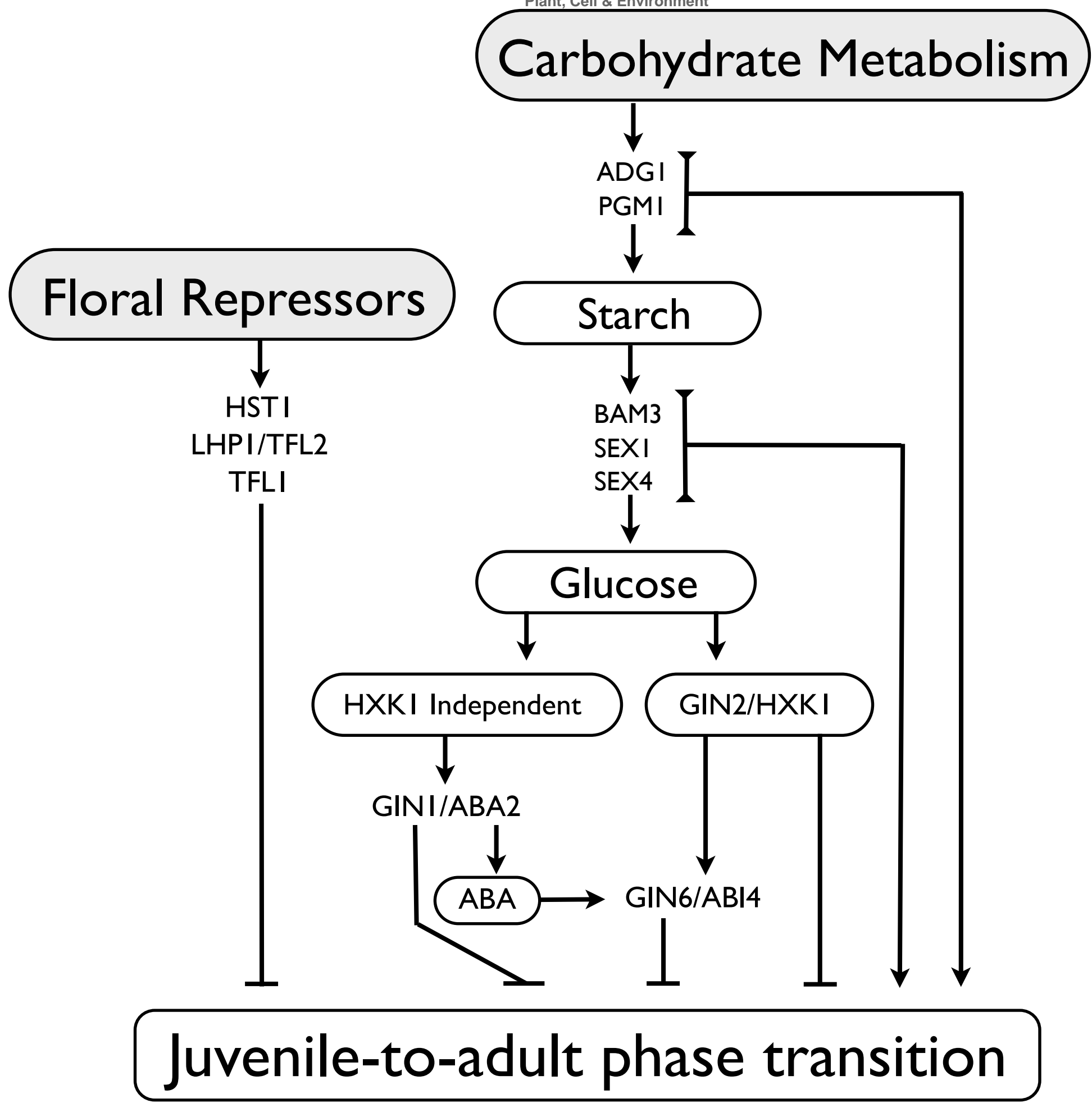

**Figure 5**

**Table 1.** Estimation of the juvenile phase lengths of the *Arabidopsis* genotypes based on attainment of floral competence.

| Genotype | Function | Juvenile phase length (days) |
|---|---|---|
| L*er*-0 | WT | 5.4 (± 0.3) |
| *tfl1* | Floral repressor | 0.3 (± 0.2) |
| *hst1* | Floral repressor | 1.2 (± 0.2) |
| *gin1-1* (*aba2*) | Sugar sensor | 3.0 (± 0.2) |
| *gin2-1* (*hxk1*) | Sugar sensor | 2.5 (± 0.3) |
| Col-0 | WT | 2.0 (± 0.2) |
| *gin6* (*abi4*) | Sugar sensor/ABA | 1.7 (± 0.2) |
| *pgm1* | Starch deficient | 5.0 (± 0.2) |
| *adg1-1* | Starch deficient | 5.9 (± 0.4) |
| *sex1* | Starch excess | 3.5 (± 0.2) |
| *sex4* | Starch excess | 5.0 (± 0.3) |
| *bam3* | Starch excess | 4.0 (± 0.2) |
| Ws-4 | WT | 7.6 (± 0.4) |
| *lhp1* (*tfl2*) | Floral repressor | 1.0 (± 0.2) |

The juvenile phase length is expressed as number of days from 50% of seedling emergence. The length of juvenile phase is estimated based on attainment of floral competence. Standard error of mean (SEM) of the estimated juvenile phase length indicated in parenthesis.

**Table 1**

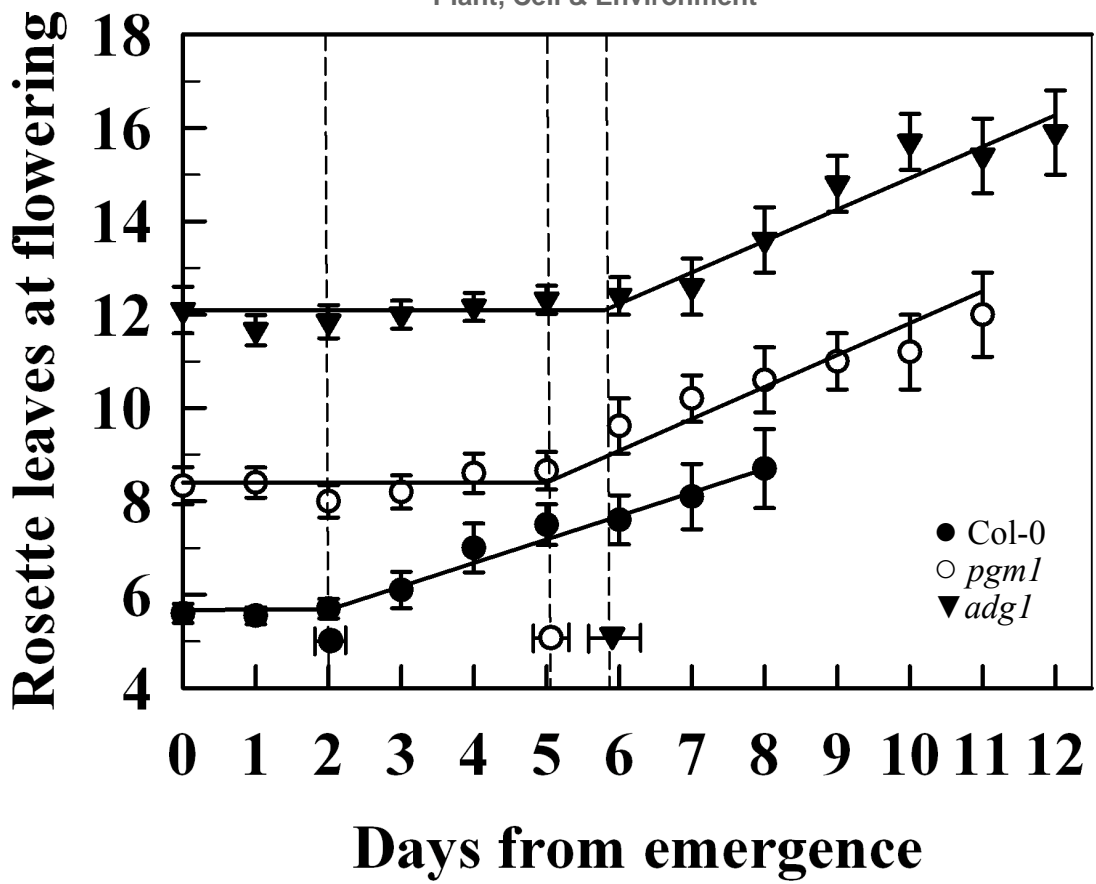

**A**

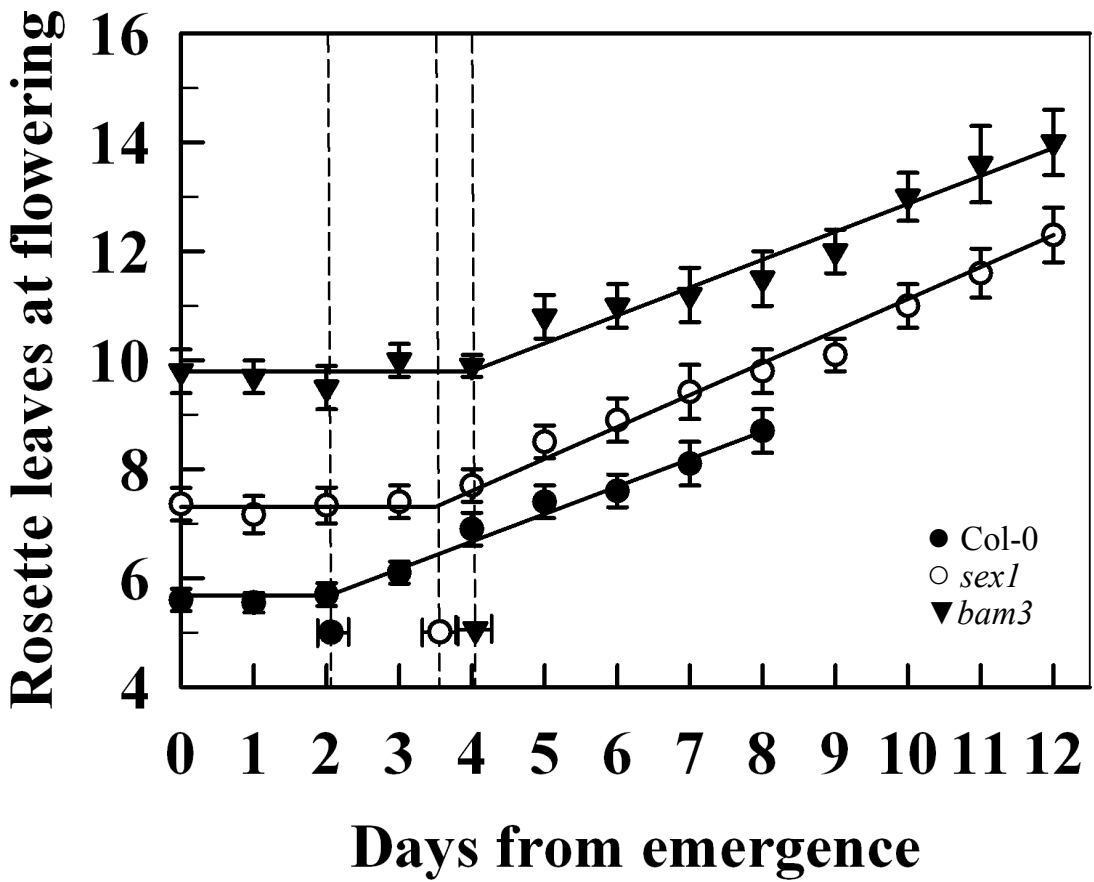

**B**

**Supplementary Fig. S1** Estimation of the juvenile phase length of *Arabidopsis* mutants involved in starch anabolism and catabolism-related events based on attainment of floral competence.

(A) Estimation of the juvenile phase length of Col-0 (●), *pgm1* (○) and *adg1-1* (▼) mutant genotypes. (B) Estimation of the juvenile phase length of Col-0 (●), *sex1-1* (○) and *bam3* (▼) mutant genotypes. Plants were transferred from SDs to LDs at time intervals shown in x-axes. Points and vertical error bars denote the mean and SEM of leaves at flowering of replicate plants transferred on each occasion. The dashed line delimits the length of juvenile phase. Horizontal error bars denote the SEM of the estimated juvenile phase length.

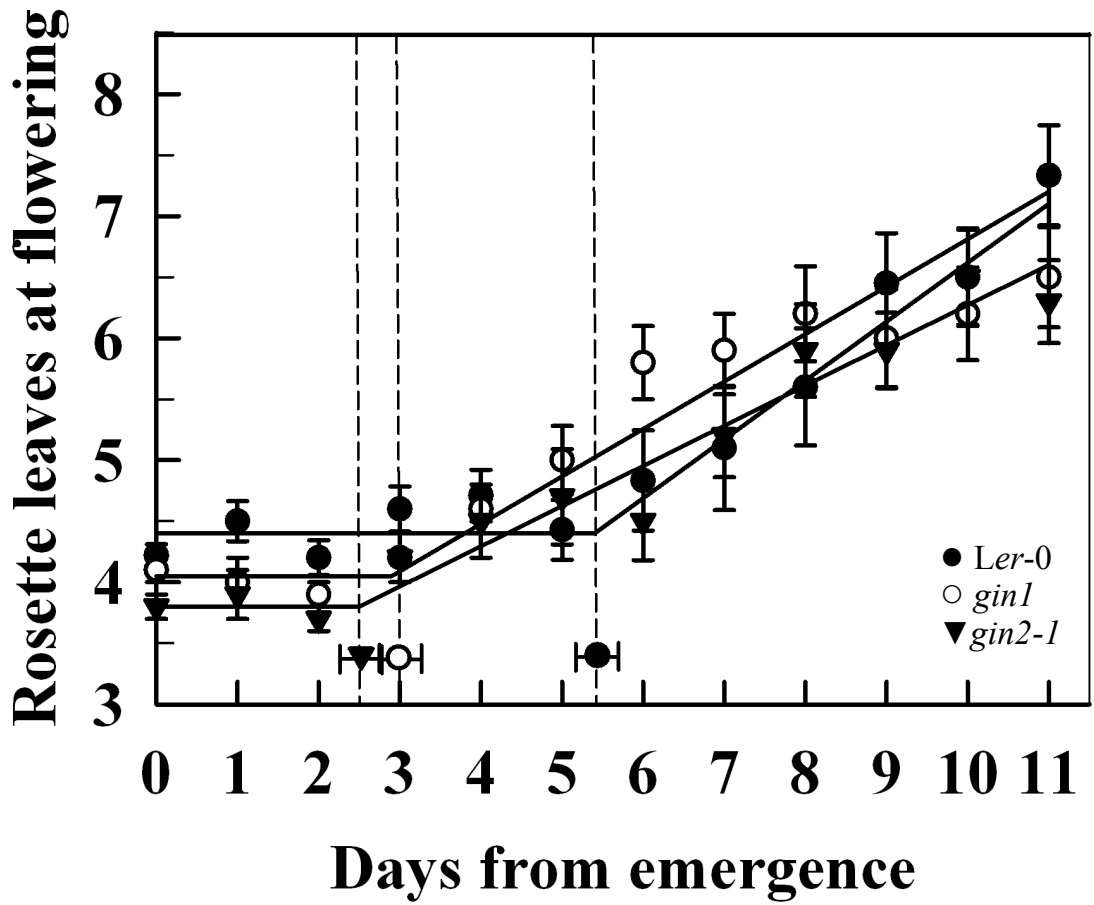

**A**

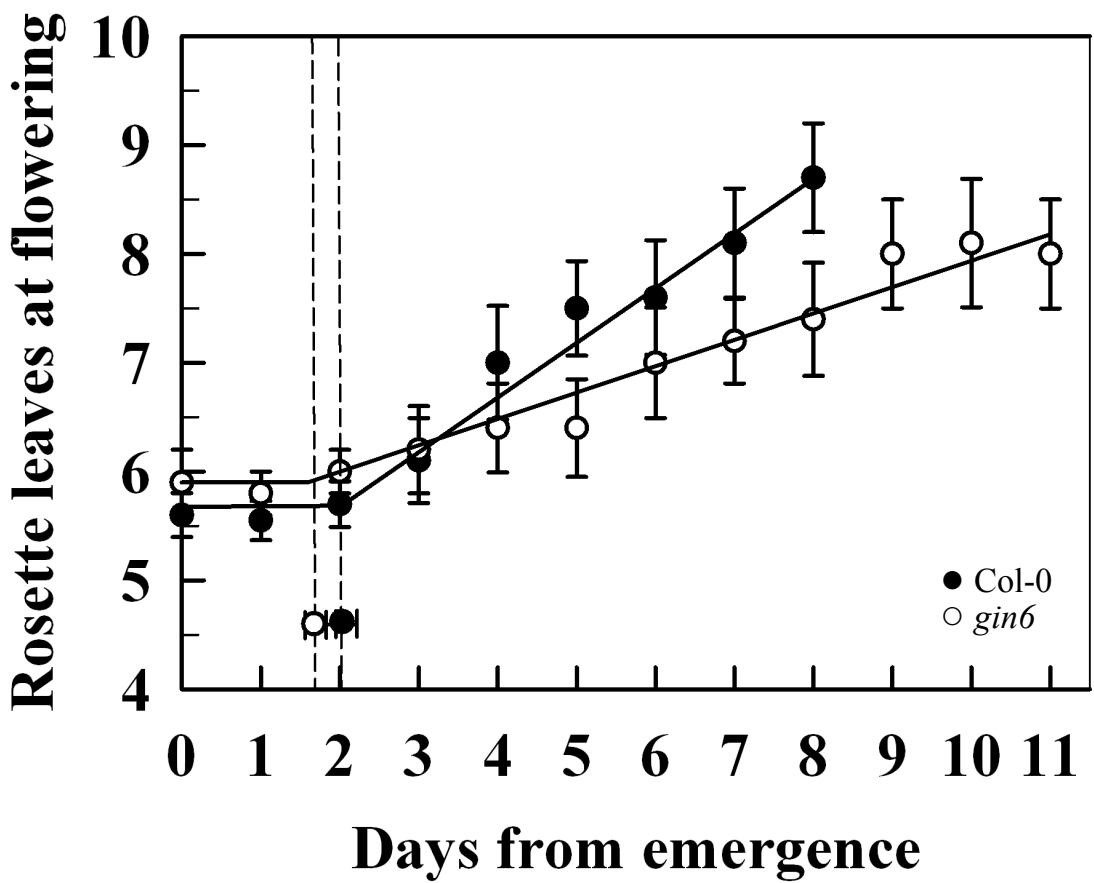

**B**

**Supplementary Fig. S2** Estimation of the juvenile phase length of glucose insensitive mutants of *Arabidopsis* based on attainment of floral competence.

(A) Estimation of the juvenile phase length of L*er*-0 (●), *gin1-1* (○) and *gin2-1* (▼). (B) Estimation of the juvenile phase length of Col-0 (●) and *gin6* (○) mutant genotypes. Plants were transferred from SDs to LDs at time intervals shown in x-axes. Points and vertical error bars denote the mean and SEM of leaves at flowering of replicate plants transferred on each occasion. The dashed line delimits the length of juvenile phase. Horizontal error bars denote the SEM of the estimated juvenile phase length.

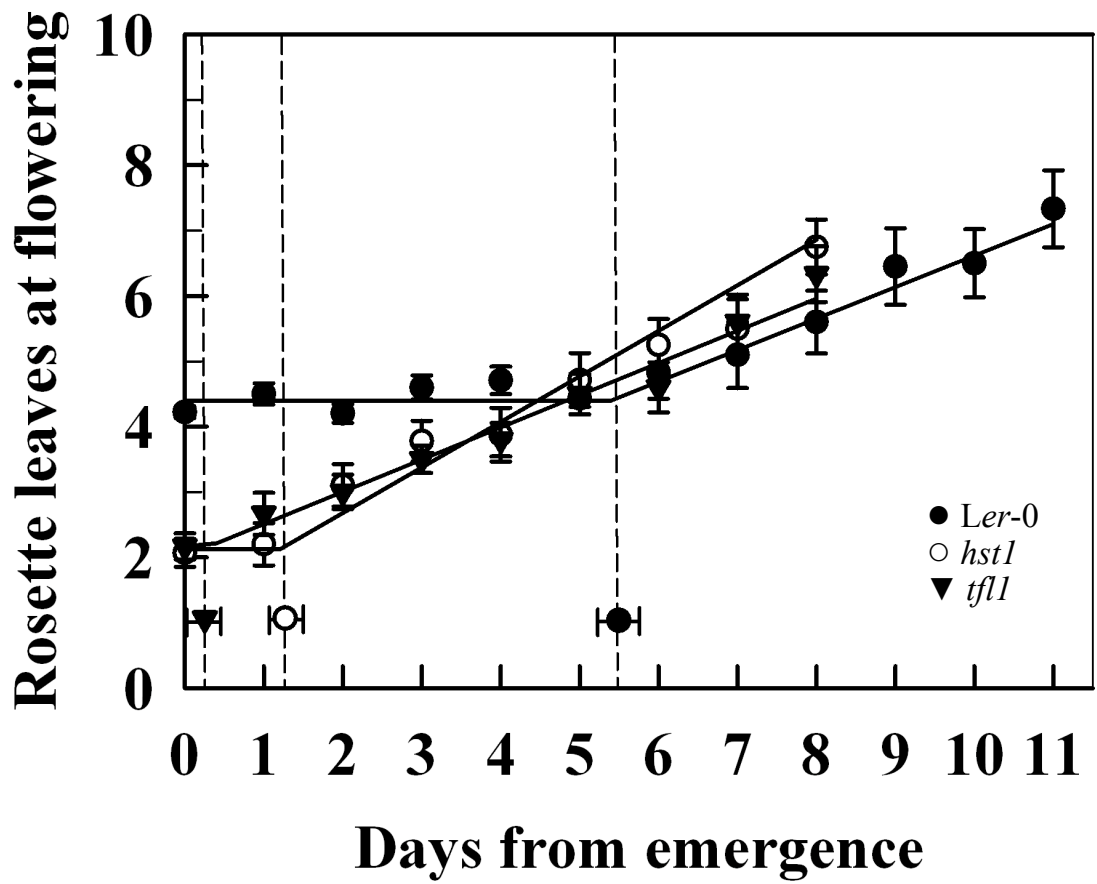

**A**

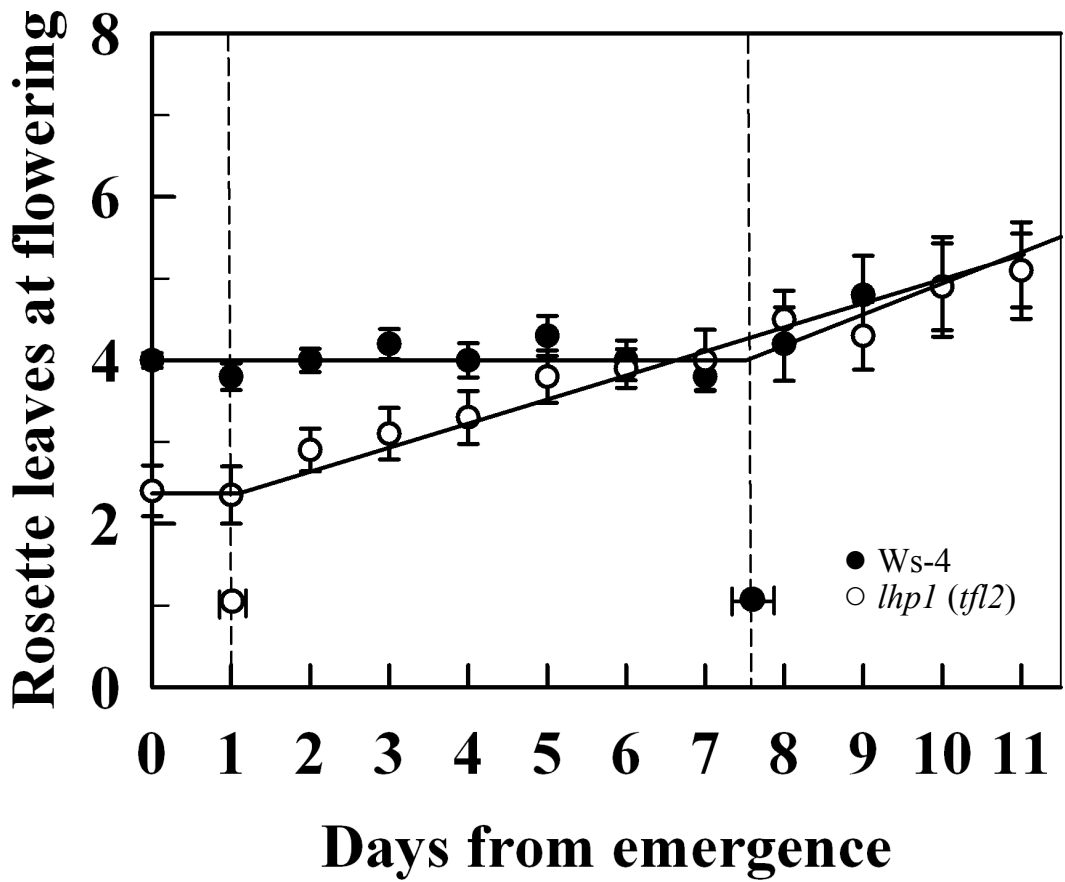

**B**

**Supplementary Fig. S3** Estimation of the juvenile phase length of floral repressors mutants of *Arabidopsis* based on attainment of floral competence.

(A) Estimation of the juvenile phase length of L*er*-0 (●), *hst1* (○) and *tfl1* (▼). (B) Estimation of the juvenile phase length of Ws-4 (●) and *lhp1* (*tfl2*) (○) mutant genotypes. Plants were transferred from SDs to LDs at time intervals shown in x-axes. Points and vertical error bars denote the mean and SEM of leaves at flowering of replicate plants transferred on each occasion. The dashed line delimits the length of juvenile phase. Horizontal error bars denote the SEM of the estimated juvenile phase length.

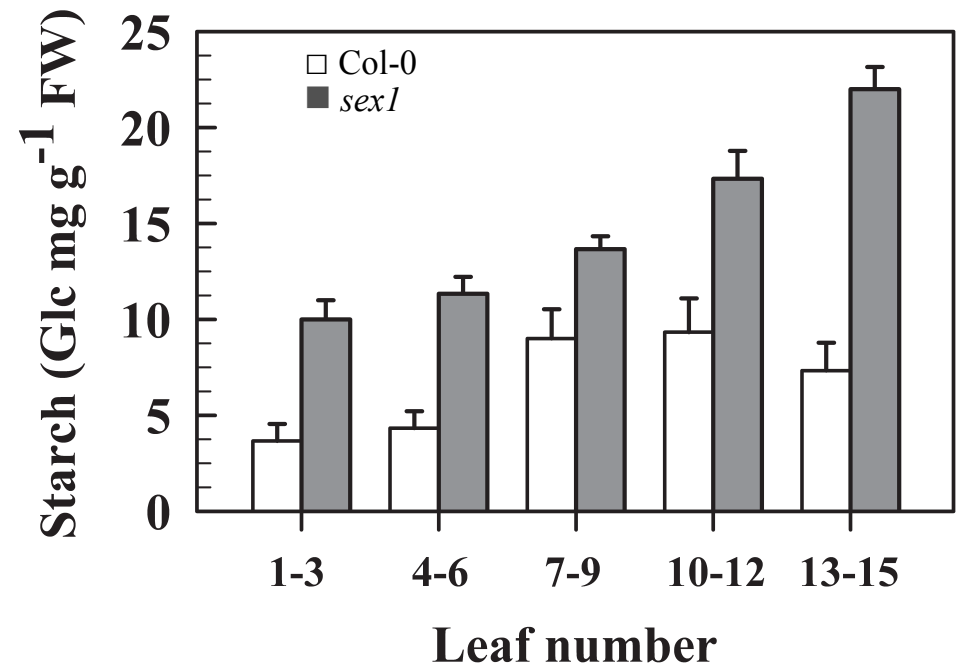

**Supplementary Fig. S4** Starch content of leaves of different developmental ages of Col-0 WT and *sex1* mutant.

Starch content of leaves of different developmental ages from Col-0 WT (□) and *sex1* (■) mutant of *Arabidopsis*. Material was collected at ZT 7 under SD conditions, at developmental stages shown in x-axis. Leaf 1 and leaf 15 denote the youngest and oldest leaves, respectively. All results are the SEM of three biological replicates.

**Supplementary Table S1:** Information on the *Arabidopsis* genotypes used in this study.

| Genotype | Ecotype | Function | EASC ID | Mutagen | Reference |
|---|---|---|---|---|---|
| L*er*-0 | | WT | NW20 | | (Koornneef *et al.* 1991) |
| Col-0 | | WT | N1092 | | (Koornneef *et al.* 1991) |
| Ws-4 | | WT | N5390 | | (Koornneef *et al.* 1991) |
| *tfl1* | L*er* | Floral repressor | N3091 | EMS[1] | (Ratcliffe *et al.* 1999) |
| *hst1* | L*er* | Floral repressor | N3811 | DEB[2] | (Telfer & Poethig 1998) |
| *gin1-1* | L*er* | Sugar sensor | Sheen, J. | EMS | (Zhou *et al.* 1998) |
| *gin2-1* | L*er* | Sugar sensor | Sheen, J. | EMS | (Zhou *et al.* 1998) |
| *gin6* | Col-0 | Sugar sensor/ABA | N122591 | Tn[3] | (Finkelstein *et al.* 1998) |
| *pgm1* | Col-0 | Starch deficient | N210 | EMS | (Caspar *et al.* 1985) |
| *adg1-1* | Col-0 | Starch deficient | N3094 | EMS | (Wang *et al.* 1998) |
| *sex1-1* | Col-0 | Starch excess | N3093 | EMS | (Yu *et al.* 2001) |
| *sex4* | Col-0 | Starch excess | Zeeman, S. | X[4] | (Zeeman *et al.* 1998) |
| *bam3* | Col-0 | Starch excess | Zeeman, S. | TILLING[5] | (Lao *et al.* 1999) |
| *lhp1* | Ws-4 | Floral repressor | Gaudin, V. | T-DNA[6] | (Gaudin *et al.* 2001) |

Names in EASC ID column indicate the donor other than EASC. [1]EMS: Ethylmethanesulfonate, [2]DEB: diepoxybutane, [3]Tn: Transposon, [4]X: X-rays, [5]TILLING: Targeting Induced Local Lesions in Genomes, [6]T-DNA: Transferred DNA, EASC: European *Arabidopsis* Stock Centre.